\documentclass[journal,10pt]{IEEEtran}
\usepackage{amsfonts}
\usepackage{cite}
\usepackage{graphicx}
\usepackage{float}
\usepackage{multicol}
\usepackage{bm}
\usepackage{upgreek}
\usepackage{color}
\usepackage{amssymb,amsmath}
\usepackage{balance}
\usepackage{booktabs}
\usepackage{stfloats}
\usepackage{algorithm,algorithmic,float}
\usepackage{algorithm}
\usepackage{algorithmic}
\usepackage{float}
\usepackage{lipsum}
\usepackage{multirow}
\makeatletter
\renewcommand{\maketag@@@}[1]{\hbox{\m@th\normalsize\normalfont#1}}%
\makeatother
\begin{document}

\title{Energy Efficiency Maximization in IRS-Aided Cell-Free Massive MIMO System}

\author{Si-Nian Jin, \emph{Member, IEEE}, Dian-Wu Yue, \emph{Senior Member, IEEE}, Yi-Ling Chen, Qing Hu, \emph{Member, IEEE}
\thanks{S.-N. Jin, D.-W. Yue and Q. Hu are with the College of Information Science and Technology, Dalian Maritime University, Dalian 116026, China (e-mail: jinsinian@dlmu.edu.cn; dwyue@dlmu.edu.cn; hq0518@dlmu.edu.cn).

Y.-L. Chen is with the School of Intelligence and Electronic Engineering, Dalian Neusoft University of Information, Dalian 116023, China (e-mail: chenyiling@neusoft.edu.cn).}}

\maketitle

\begin{abstract}
In this paper, we consider an intelligent reflecting surface (IRS)-aided cell-free massive multiple-input multiple-output system, where the beamforming at access points and the phase shifts at IRSs are jointly optimized to maximize energy efficiency (EE). To solve EE maximization problem, we propose an iterative optimization algorithm by using quadratic transform and Lagrangian dual transform to find the optimum beamforming and phase shifts. However, the proposed algorithm suffers from high computational complexity, which hinders its application in some practical scenarios. Responding to this, we further propose a deep learning based approach for joint beamforming and phase shifts design. Specifically, a two-stage deep neural network is trained offline using the unsupervised learning manner, which is then deployed online for the predictions of beamforming and phase shifts. Simulation results show that compared with the iterative optimization algorithm and the genetic algorithm, the unsupervised learning based approach has higher EE performance and lower running time.
\end{abstract}

\IEEEpeerreviewmaketitle
\vspace{-0.4cm}
\section{Introduction}
Recently, cell-free massive multiple-input multiple-output (MIMO) has emerged as a promising technology to effectively alleviate inter-cell interference \cite{Ngo17}. In this system, a large number of distributed access points (APs) are linked to the central processing unit (CPU) through the backhaul link and provide services to all the users without cell boundaries. However, the large-scale deployment of APs will bring some problems, such as high deployment cost and power consumption. To address these difficulties, a promising technique called intelligent reflecting surface (IRS) has been proposed as a cost-efficient and energy-efficient solution \cite{Wu19,Wu18}. Specifically, IRS consists of a large number of passive low-cost reflecting elements, each of which can reflect the electromagnetic incident signals to any directions, constructing a favorable propagation environment.

Inspired by IRS, IRS-aided cell-free massive MIMO system has received a lot of attention, where spectral efficiency was extensively studied \cite{Jin22,ZZhang21}. Besides, energy efficiency (EE) as another key performance metric, some existing works also investigated the EE of this system \cite{Zhang21,Le21,Siddiqi22,Liu21}. For instance, the researchers in \cite{Zhang21,Le21,Siddiqi22} proposed some iterative optimization algorithms to solve total EE maximization problem. Furthermore, in order to maintain fairness among users, an alternate optimization algorithm was proposed to enhance the EE of the worst user \cite{Liu21}. However, all of the above-mentioned iterative optimization algorithms have high computational complexity. This makes these algorithms unsuitable for some practical scenarios, such as high mobility communication scenarios. To overcome the drawback of conventional algorithms, in the context of IRS-aided communication systems, deep learning (DL) based on approach was recently proposed to address related issues, mainly focusing on the spectral efficiency optimization \cite{Song21,Huang20}. The significant advantage of DL approach is that it can provide ideal predictions with low computational complexity based on trained neural network. Although the DL approach has many advantages, it is still a blank for research work using DL approach to solve EE maximization problem in IRS-aided cell-free massive MIMO system.

To fill this gap, this paper investigates the EE maximization problem by jointly optimizing the beamforming at APs and the phase shifts at IRSs. First, since the EE maximization problem in this paper is partially different from the problem in \cite{Zhang21,Le21,Siddiqi22,Liu21}, we develop a novel iterative optimization algorithm to solve the corresponding problem. Then, an unsupervised learning based approach is proposed to tackle the EE maximization problem. Specifically, we model a two-stage deep neural network (DNN) and design a reasonable loss function, and train this DNN in an unsupervised manner to learn the optimal beamforming and phase shifts. At last, simulation results show that compared with the traditional genetic algorithm (GA) and the proposed iterative optimization algorithm, the unsupervised learning based approach can achieve better EE performance with extremely low running time.

The rest of this paper is organized as follows. Section II presents system model and problem formulation. Section III and Section IV propose an iterative optimization algorithm and an unsupervised learning based approach to solve EE maximization problem, respectively. Numerical results are discussed in Section V, and Section VI concludes this paper.

\emph{Notations:}  ${\left( {\cdot} \right)^*}$, ${\left( {\cdot} \right)^{\rm T}}$, ${\left( {\cdot} \right)^{\rm H}}$, ${\rm trace}\left( {\cdot} \right)$, $\left| {\cdot} \right|$, $\left\| {\cdot} \right\|$, ${\mathop{\rm Re}\nolimits} \left\{ {\bf{\cdot}}  \right\}$ and ${\rm{diag}}\left\{ {\bf{\cdot}} \right\}$ denote conjugate, transpose, conjugate-transpose, trace, absolute value, Euclidean norm, real part and diagonal operation, respectively. $\mathbb{C}$ and $\mathbb{R}$ denote the sets of complex and real numbers, respectively. $\mathcal{CN}\left( {0, \sigma^2} \right)$ denotes a circularly symmetric complex Gaussian random variable with mean 0 and covariance $\sigma^2$. ${\left[ {\bf{X}} \right]_{mn}}$ denotes the $\left( {m,n} \right){\mathop{\rm th}\nolimits}$  element of matrix ${\bf{X}}$. ${\left[ x \right]^ + }$ denotes the operation of $\max \left\{ {0,x} \right\}$.

\section{\small SYSTEM MODEL AND PROBLEM FORMULATION}
\begin{figure}[h]
\setlength{\abovecaptionskip}{0.cm}
\setlength{\belowcaptionskip}{-0.cm}
\centering
\includegraphics[scale=0.47]{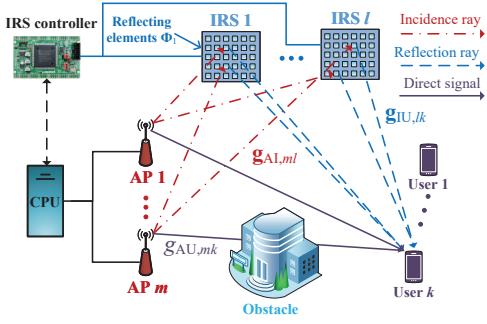}
\caption{The downlink IRS-aided cell-free massive MIMO system.}\label{fig1}
\end{figure}
This paper considers the downlink IRS-aided cell-free massive MIMO system, as shown in Fig. 1. For this system, through the instructions of the CPU, $M$ single-antenna APs are deployed to serve $K$ single-antenna users with the help of $L$ IRSs, where each IRS is equipped with $N$ reflecting elements and the total number of reflecting elements at all IRSs is $I=LN$. Let $\mathcal{M} = \left\{ {1, \cdots ,M} \right\}$, $\mathcal{K} = \left\{ {1, \cdots ,K} \right\}$ and $\mathcal{I} = \left\{ {1, \cdots ,I} \right\}$ as the index sets of the APs, the users and the total elements of all IRSs, respectively.
\vspace{-0.3cm}
\subsection{Transmission Model}
The notation ${{g}_{{\rm AU},mk}} \in {\mathbb{C}^{1 \times 1}}$, ${{\bf{g}}_{{\rm AI},ml}} \in {\mathbb{C}^{N \times 1}}$, and ${{\bf{g}}_{{\rm IU},lk}^{\rm H}} \in {\mathbb{C}^{1 \times N}}$ denote the channel from AP $m$ to user $k$, the channel from AP $m$ to IRS $l$, and the channel from IRS $l$ to user $k$, respectively. For convenience, we define ${{\bf{g}}_{{\mathop{\rm AU}\nolimits} ,k}^{\rm H}} = \left[ {{g_{{\mathop{\rm AU}\nolimits} ,1k}}, \cdots,{g_{{\mathop{\rm AU}\nolimits} ,Mk}}} \right] \in {\mathbb{C}^{1 \times M}}$ and ${{\bf{G}}_{{\mathop{\rm AI}\nolimits} ,l}} = \left[ {{{\bf{g}}_{{\mathop{\rm AI}\nolimits} ,1l}}, \cdots ,{{\bf{g}}_{{\mathop{\rm AI}\nolimits} ,Ml}}} \right] \in {\mathbb{C}^{N \times M}}$ as the channel vector from all APs to user $k$ and the channel matrix from all APs to IRS $l$, respectively. In addition, by using the channel estimation methods \cite{Wang20}, all channel state information is assumed to be perfectly obtained by CPU. Due to the existence of IRS, there are some cascaded channels between the APs and the users, we define ${\bf{g}}_{{\mathop{\rm AIU}\nolimits} ,k}^{\rm H} \in {\mathbb{C}^{1 \times M}}$ as the composite AP-IRS-user channel from all APs to user $k$ via all IRSs, which is given as
\begin{equation}\label{1}
  {{\bf{g}}_{{\mathop{\rm AIU}\nolimits} ,k}^{\rm H}} = \sum\nolimits_{l = 1}^L {{\bf{g}}_{{\mathop{\rm IU}\nolimits} ,lk}^{\mathop{\rm H}\nolimits} {{\bf{\Phi }}_l}{{\bf{G}}_{{\mathop{\rm A}\nolimits} {\mathop{\rm I}\nolimits} ,l}}}  = {{\bf{v}}^{\mathop{\rm H}\nolimits} }{{\bf{G}}_{{\mathop{\rm AIU}\nolimits} ,k}},\;\;k \in \mathcal{K},
\end{equation}
where the phase shift matrix of IRS $l$ can be optimized through a diagonal matrix ${{\bf{\Phi }}_l} = {\mathop{\rm diag}\nolimits} \left\{ {{{\mathop{\rm e}\nolimits} ^{j{\theta _{l1}}}}, \cdots ,{{\mathop{\rm e}\nolimits} ^{j{\theta _{lN}}}}} \right\}$ with ${\theta _{ln}} \in \left[ {0,2\pi } \right]$, and other notations are defined as
\begin{equation}\label{2}
{{\bf{v}}^{\mathop{\rm H}\nolimits} } = \left[ {{{\rm{e}}^{j{\theta _{11}}}}, \cdots ,{{\rm{e}}^{j{\theta _{1N}}}}, \cdots ,{{\rm{e}}^{j{\theta _{L1}}}}, \cdots ,{{\rm{e}}^{j{\theta _{LN}}}}} \right] \in {\mathbb{C}^{1 \times I}},
\end{equation}
\begin{equation}\label{3}
{{\bf{G}}_{{\mathop{\rm AIU}\nolimits} ,lk}} = {\rm{diag}}\left\{ {{\bf{g}}_{{\mathop{\rm IU}\nolimits} ,lk}^{\mathop{\rm H}\nolimits} } \right\}{{\bf{G}}_{{\mathop{\rm A}\nolimits} {\mathop{\rm I}\nolimits} ,l}}\in {\mathbb{C}^{N \times M}},
\end{equation}
\begin{equation}\label{4}
{{\bf{G}}_{{\mathop{\rm AIU}\nolimits} ,k}} = {\left[ {{\bf{G}}_{{\mathop{\rm AIU}\nolimits} ,1k}^{\mathop{\rm T}\nolimits} , \cdots ,{\bf{G}}_{{\mathop{\rm AIU}\nolimits} ,Lk}^{\mathop{\rm T}\nolimits} } \right]^{\mathop{\rm T}\nolimits} }\in {\mathbb{C}^{I \times M}}.
\end{equation}

Based on the above system model, the received signal of user $k$ can be expressed as
\begin{equation}\label{5}
{r_k} = \left( {{{\bf{v}}^{\mathop{\rm H}\nolimits} }{{\bf{G}}_{{\mathop{\rm AIU}\nolimits} ,k}} + {{\bf{g}}_{{\mathop{\rm AU}\nolimits} ,k}^{\rm H}}} \right)\sum\nolimits_{j = 1}^K {{{\bf{w}}_j}{x_j}}  + {n_k},\;\;k \in \mathcal{K},
\end{equation}
where ${\bf{W}} \in {\mathbb{C}^{M \times K}}$ represents the beamforming matrix at all APs, which its $m$th row and $k$th column element is ${\left[ {\bf{W}} \right]_{mk}} = {w_{mk}}$. Thus, ${{\bf{w}}_j} = {\left[ {{w_{1j}}, \cdots ,{w_{Mj}}} \right]^{\mathop{\rm T}\nolimits} } \in {\mathbb{C}^{M \times 1}}$ is the $j$th column of ${\bf{W}}$, representing the beamforming vector for user $j$. ${x_j}$ and ${n_k} \sim \mathcal{CN}\left( {0,{{\sigma ^2}}} \right)$ are the unit-power information symbol for user $j$ and the noise at user $k$, respectively. Based on (5), the achievable rate of user $k$ can be given by
\begin{small}
\begin{equation}\label{6}
{R_k} = {\log _2}\left( {1 + \frac{{{{\left| {\left( {{{\bf{v}}^{\mathop{\rm H}\nolimits} }{{\bf{G}}_{{\mathop{\rm AIU}\nolimits} ,k}} + {{\bf{g}}_{{\mathop{\rm AU}\nolimits} ,k}^{\rm H}}} \right){{\bf{w}}_k}} \right|}^2}}}{{\sum\limits_{j \ne k}^K {{{\left| {\left( {{{\bf{v}}^{\mathop{\rm H}\nolimits} }{{\bf{G}}_{{\mathop{\rm AIU}\nolimits} ,k}} + {{\bf{g}}_{{\mathop{\rm AU}\nolimits} ,k}^{\rm H}}} \right){{\bf{w}}_j}} \right|}^2}}  + {\sigma ^2}}}} \right).
\end{equation}
\end{small}
\vspace{-0.9cm}
\subsection{Optimization Problem Formulation}
To implement the design of EE optimization algorithm, we define the total power consumption of system as
\begin{equation}\label{7}
{P_{{\mathop{\rm total}\nolimits} }} = \sum\limits_{k = 1}^K {\alpha {\bf{w}}_k^{\mathop{\rm H}\nolimits} {{\bf{w}}_k}}  + {P_{{\mathop{\rm fix}\nolimits} }},
\end{equation}
where $\sum\nolimits_{k = 1}^K {\alpha {\bf{w}}_k^{\mathop{\rm H}\nolimits} {{\bf{w}}_k}}$ is the transmit power of all APs and $\alpha  = {\upsilon ^{ - 1}}$ with $\upsilon  \in \left( {0,1} \right]$ being the power amplifier efficiency. Furthermore, ${P_{{\mathop{\rm fix}\nolimits} }} = M{P_{{\mathop{\rm AP}\nolimits} }} + K{P_{{\rm{User}}}} + I{P_{{\mathop{\rm IRS}\nolimits} }}$ denotes the static power consumption, where ${P_{{\rm{AP}}}}$, ${P_{{\rm{User}}}}$ and ${P_{{\mathop{\rm IRS}\nolimits} }}$  are the circuit power consumption of each AP, each user and each reflecting element of IRS, respectively. Thus, EE is given by
\begin{equation}\label{8}
{\mathop{\rm EE}\nolimits}  = \frac{{B\sum\nolimits_{k = 1}^K {{R_k}} }}{{\sum\nolimits_{k = 1}^K {\alpha {\bf{w}}_k^{\mathop{\rm H}\nolimits} {{\bf{w}}_k}}  + {P_{{\mathop{\rm fix}\nolimits} }}}},
\end{equation}
where $B$ is the bandwidth of the channel.

Based on (8), the problem of maximizing EE can be formulated as
\setcounter{equation}{8}
\begin{align}
\mathop {\max }\limits_{{\bf{W}},{\bf{v}}} \;\;&{\mathop{\rm EE}\nolimits}\tag{9a}\\
\;{\mathop{\rm s.t.}\nolimits} \;\;\;&{R_k} \ge {R_{\min }},\;\;\;\;\;\;\;\;\;\;\;\;\;\;k \in \mathcal{K},\tag{9b}\\
\;\;\;\;\;\;\;\;\;&{\left\| {{{{\bf{\bar w}}}_m}} \right\|^2} \le {P_{\max }},\;\;\;\;\;\;\;m \in \mathcal{M},\tag{9c}\\
\;\;\;\;\;\;\;\;\;&{\left| {{{\left[ {\bf{v}} \right]}_i}} \right|^2} = 1,\;\;\;\;\;\;\;\;\;\;\;\;\;\;\;\;i \in \mathcal{I}.\tag{9d}
\end{align}

The constraint in (9b) guarantees the user's rate requirements of all users, where $R_{\min}$ is the minimum data rate requirement. The constraint in (9c) is a transmit power constraint, where ${{\bf{\bar w}}_m} = \left[ {{w_{m1}}, \cdots ,{w_{mK}}} \right] \in {\mathbb{C}^{1 \times K}}$ and $P_{\max }$ are the beamforming vector at AP $m$ and the maximum transmit power for each AP, respectively. The unit modulus constraint for each reflecting element of IRS is provided in (9d). Due to the non-convexity of this problem, it is difficult to solve. 

\section{ITERATIVE OPTIMIZATION ALGORITHM}
To solve the problem of maximizing EE, we first decompose the problem (9) into two subproblems, i.e., the beamforming at APs and the phase shifts at IRSs, respectively. Then, we design the corresponding algorithm to alternately optimize them.
\subsection{Optimizing ${\bf{W}}$ for Given ${\bf{v}}$}
For a fixed ${\bf{v}}$, problem (9) can be transformed into
\setcounter{equation}{9}
\begin{align}\label{10}
\mathop {\max }\limits_{\bf{W}} \;\;&\frac{{B\sum\nolimits_{k = 1}^K {{{\log }_2}\left( {1 + \frac{{{{\left| {{{\bf{h}}_{{\mathop{\rm AU}\nolimits} ,k}^{\rm H}}{{\bf{w}}_k}} \right|}^2}}}{{\sum\nolimits_{j \ne k}^K {{{\left| {{{\bf{h}}_{{\mathop{\rm AU}\nolimits} ,k}^{\rm H}}{{\bf{w}}_j}} \right|}^2}}  + {\sigma ^2}}}} \right)} }}{{\sum\nolimits_{k = 1}^K {\alpha {\bf{w}}_k^{\mathop{\rm H}\nolimits} {{\bf{w}}_k}}  + {P_{{\mathop{\rm fix}\nolimits} }}}}\tag{10a}\\
\;{\mathop{\rm s.t.}\nolimits} \;\;&\left\| {{\bf{h}}_{{\mathop{\rm AU}\nolimits} ,k}^{\mathop{\rm H}\nolimits} {{\bf{W}}_k},{\sigma}} \right\| \le \frac{{{\bf{h}}_{{\mathop{\rm AU}\nolimits} ,k}^{\mathop{\rm H}\nolimits} {{\bf{w}}_k}}}{{\sqrt {{2^{{R_{\min }}}} - 1} }},\;\;\;\;k \in \mathcal{K},\tag{10b}\\
\;\;\;\;\;\;\;\;&{\left\| {{{{\bf{\bar w}}}_m}} \right\|} \le {\sqrt {P_{\max }}},\;\;\;m \in \mathcal{M},\tag{10c}
\end{align}
where ${{\bf{W}}_k} = \left[ {{{\bf{w}}_1}, \cdots ,{{\bf{w}}_{k - 1}},{{\bf{w}}_{k + 1}}, \cdots ,{{\bf{w}}_K}} \right] \in {\mathbb{C}^{M \times \left( {K - 1} \right)}}$ and ${\bf{h}}_{{\mathop{\rm AU}\nolimits} ,k}^{\mathop{\rm H}\nolimits}  = {{\bf{v}}^{\mathop{\rm H}\nolimits} }{{\bf{G}}_{{\mathop{\rm AIU}\nolimits} ,k}} + {{\bf{g}}_{{\mathop{\rm AU}\nolimits} ,k}^{\rm H}}$ represents an aggregated channel between all APs and user $k$.

In problem (10), the constraints of (10b) and (10c) are convex, but the objective function of (10a) is non-convex. In order to convert problem (10) into convex problem, we apply the quadratic transform \cite{Shen18} to the objective function (10a), so problem (10) can be transformed into
\setcounter{equation}{10}
\begin{align}
\mathop {\max }\limits_{{\bf{W}},{\bf{y}},z} \;\;&{f_1}\left( {{\bf{W}},{\bf{y}},z} \right)\tag{11a}\\
\;{\mathop{\rm s.t.}\nolimits} \;\;\;&\left( {10{\rm{b}}} \right),\left( {10{\mathop{\rm c}\nolimits} } \right)\tag{11b},
\end{align}
where ${\bf{y}}$ refers to a collection of variables $\left\{ {{y_1}{\rm{,}} \cdots {\rm{,}}{y_K}} \right\}$ and
\setcounter{equation}{11}
\begin{equation}\label{12}
{y_k} = \frac{{{\bf{h}}_{{\mathop{\rm AU}\nolimits} ,k}^{\mathop{\rm H}\nolimits} {{\bf{w}}_k}}}{{\sum\nolimits_{j \ne k}^K {{{\left| {{\bf{h}}_{{\mathop{\rm AU}\nolimits} ,k}^{\mathop{\rm H}\nolimits} {{\bf{w}}_j}} \right|}^2}}  + {\sigma ^2}}},\;k \in \mathcal{K},
\end{equation}
\begin{equation}\label{13}
{z} = \frac{{\sqrt {\sum\nolimits_{k = 1}^K {{{\log }_2}\left( {1 + \frac{{{{\left| {{{\bf{h}}_{{\mathop{\rm AU}\nolimits} ,k}^{\rm H}}{{\bf{w}}_k}} \right|}^2}}}{{\sum\nolimits_{j \ne k}^K {{{\left| {{{\bf{h}}_{{\mathop{\rm AU}\nolimits} ,k}^{\rm H}}{{\bf{w}}_j}} \right|}^2}}  + {\sigma ^2}}}} \right)} } }}{{\alpha \sum\nolimits_{k = 1}^K {{\bf{w}}_k^{\mathop{\rm H}\nolimits} {{\bf{w}}_k}}  + {P_{{\mathop{\rm fix}\nolimits} }}}},
\end{equation}
\begin{equation}\label{14}
\begin{array}{l}
{f_1}\left( {{\bf{W}},{\bf{y}},z} \right) =  - {z^2}\left( {\alpha \sum\limits_{k = 1}^K {{\bf{w}}_k^{\mathop{\rm H}\nolimits} {{\bf{w}}_k}}  + {P_{{\mathop{\rm fix}\nolimits} }}} \right) + \\
2z\sqrt {\sum\limits_{k = 1}^K {{{\log }_2}\left( \begin{array}{l}
1 + 2{\mathop{\rm Re}\nolimits} \left\{ {y_k^ * {\bf{h}}_{{\mathop{\rm AU}\nolimits} ,k}^{\mathop{\rm H}\nolimits} {{\bf{w}}_k}} \right\} -\\
 {\left| {{y_k}} \right|^2}\left( {\sum\limits_{j \ne k}^K {{{\left| {{\bf{h}}_{{\mathop{\rm AU}\nolimits} ,k}^{\mathop{\rm H}\nolimits} {{\bf{w}}_j}} \right|}^2}}  + {\sigma ^2}} \right)
\end{array} \right)} }
\end{array}.
\end{equation}

When the auxiliary variables ${\bf{y}}$ and $z$ are fixed, problem (11) is a convex problem with respect to ${\bf{W}}$, and therefore the optimal ${\bf{W}}$ can be found by using the modern optimization solver (e.g. Mosek \cite{Mosek}). By solving problem (11), the suboptimal solution of problem (10) can also be obtained correspondingly.

\subsection{Optimizing ${\bf{v}}$ for Given ${\bf{W}}$}
For a fixed ${\bf{W}}$, problem (9) can be equivalent to the problem of maximizing sum rate as follows
\setcounter{equation}{14}
\begin{align}
\mathop {\max }\limits_{\bf{v}} \;&\sum\limits_{k = 1}^K {{{\log }_2}\left( {1 + \frac{{{{\left| {{{\bf{v}}^{\mathop{\rm H}\nolimits} }{{\bf{a}}_{kk}} + {b_{kk}}} \right|}^2}}}{{\sum\nolimits_{j \ne k}^K {{{\left| {{{\bf{v}}^{\mathop{\rm H}\nolimits} }{{\bf{a}}_{kj}} + {b_{kj}}} \right|}^2}}  + {\sigma ^2}}}} \right)} \tag{15a}\\
\;{\mathop{\rm s.t.}\nolimits} \;&\sum\limits_{j \ne k}^K {{{\left| {{{\bf{v}}^{\mathop{\rm H}\nolimits} }{{\bf{a}}_{kj}} + {b_{kj}}} \right|}^2}}  + {\sigma ^2} \le \frac{{{{\left| {{{\bf{v}}^{\mathop{\rm H}\nolimits} }{{\bf{a}}_{kk}} + {b_{kk}}} \right|}^2}}}{{{2^{{R_{\min }}}} - 1}},\;k \in \mathcal{K},\tag{15b}\\
&{\left| {{{\left[ {\bf{v}} \right]}_i}} \right|^2} = 1,\;\;\;\;\;\;\;i \in \mathcal{I},\tag{15c}
\end{align}
where ${{\bf{a}}_{kj}} = {{\bf{G}}_{{\mathop{\rm AIU}\nolimits} ,k}}{{\bf{w}}_j}$ and ${b_{kj}} = {{\bf{g}}_{{\mathop{\rm AU}\nolimits} ,k}^{\rm H}}{{\bf{w}}_j}$ for $k,j \in \mathcal{K}$.

Since the problem (15a) is non-convex, we apply the quadratic transform \cite{Shen18} and the Lagrangian dual transform \cite{KShen18} to (15a), so problem (15) can be transformed into problem (16) as follows
\setcounter{equation}{15}
\begin{align}
\mathop {\max }\limits_{{\bf{v}},{\boldsymbol{\gamma }},{\boldsymbol{\varepsilon }}} \;\;&{f_2}\left( {{\bf{v}},{\boldsymbol{\gamma }},{\boldsymbol{\varepsilon }}} \right)\tag{16a}\\
\;{\mathop{\rm s.t.}\nolimits} \;\;\;&\left( {15{\mathop{\rm b}\nolimits} } \right),\left( {15{\mathop{\rm c}\nolimits} } \right),\tag{16b}
\end{align}
where ${\boldsymbol{\gamma}} = \left[ {{\gamma _1}, \cdots ,{\gamma _K}} \right]$ and ${\boldsymbol{\varepsilon }}=\left[ {{\varepsilon _1}, \cdots ,{\varepsilon _K}} \right]$ are two auxiliary vector variables introduced to make the problem more tractable and
\setcounter{equation}{16}
\begin{equation}\label{17}
{\gamma _k} = \frac{{{{\left| {{{\bf{v}}^{\mathop{\rm H}\nolimits} }{{\bf{a}}_{kk}} + {b_{kk}}} \right|}^2}}}{{\sum\nolimits_{j \ne k}^K {{{\left| {{{\bf{v}}^{\mathop{\rm H}\nolimits} }{{\bf{a}}_{kj}} + {b_{kj}}} \right|}^2}}  + {\sigma ^2}}},\;k \in \mathcal{K},
\end{equation}
\begin{equation}\label{18}
{\varepsilon _k} = \frac{{\sqrt {1 + {\gamma _k}} \left( {{{\bf{v}}^{\mathop{\rm H}\nolimits} }{{\bf{a}}_{kk}} + {b_{kk}}} \right)}}{{\sum\nolimits_{j = 1}^K {{{\left| {{{\bf{v}}^{\mathop{\rm H}\nolimits} }{{\bf{a}}_{kj}} + {b_{kj}}} \right|}^2}}  + {\sigma ^2}}},\;k \in \mathcal{K},
\end{equation}
\begin{equation}\label{19}
{\bf{\Theta }} = \sum\limits_{k = 1}^K {\sum\limits_{j = 1}^K {{{\left| {{\varepsilon _k}} \right|}^2}{{\bf{a}}_{kj}}{\bf{a}}_{kj}^{\mathop{\rm H}\nolimits} } },
\end{equation}
\begin{equation}\label{20}
{\bf{u}} = \sum\limits_{k = 1}^K {\left( {\sqrt {1 + {\gamma _k}} \varepsilon _k^ * {{\bf{a}}_{kk}} - {{\left| {{\varepsilon _k}} \right|}^2}\sum\limits_{j = 1}^K {b_{kj}^ * {{\bf{a}}_{kj}}} } \right)},
\end{equation}
\begin{equation}\label{21}
{f_2}\left( {{\bf{v}},{\boldsymbol{\gamma }},{\boldsymbol{\varepsilon }}} \right) = \sum\limits_{k = 1}^K {\left( {{{\log }_2}\left( {1 + {\gamma _k}} \right) - {\gamma _k}} \right)}  - {{\bf{v}}^{\mathop{\rm H}\nolimits} }{\bf{\Theta v}} + 2{\mathop{\rm Re}\nolimits} \left\{ {{{\bf{v}}^{\mathop{\rm H}\nolimits} }{\bf{u}}} \right\}.
\end{equation}

By using semidefinite relaxation \cite{Wu18} and some matrix transformations for problem (16), the problem (16) can be transformed into
\setcounter{equation}{21}
\begin{align}
\mathop {\max }\limits_{\bf{Q},{\boldsymbol{\gamma }},{\boldsymbol{\varepsilon }}} \;\;&{\mathop{\rm trace}\nolimits} \left( {{\bf{\bar \Theta Q}}} \right)\tag{22a}\\
\;{\mathop{\rm s.t.}\nolimits} \;\;\;&\sum\limits_{j \ne k}^K {\left( {{\rm{trace}}\left( {{{\bf{C}}_{kj}}{\bf{Q}}} \right) + {{\left| {{b_{kj}}} \right|}^2}} \right)}  + {\sigma ^2} \nonumber\\
&\;\;\;\;\;\;\;\;\;\;\;\;\;\; \le \frac{{{\rm{trace}}\left( {{{\bf{C}}_{kk}}{\bf{Q}}} \right) + {{\left| {{b_{kk}}} \right|}^2}}}{{{2^{{R_{\min }}}} - 1}},\;\;\;k \in \mathcal{K},\tag{22b}\\
&\;{\left[ {\bf{Q}} \right]_{ii}} = 1,\;\;\;i = 1, \cdots ,I+1,\tag{22c}\\
&\;{\bf{Q}} \succeq 0,\tag{22d}
\end{align}
where ${\bf{\bar \Theta }} = \left[ \begin{array}{l}- {\bf{\Theta }},{\bf{u}}\\{\:{\bf{u}}^{\mathop{\rm H}\nolimits} },\;0\end{array} \right]$ and ${{\bf{C}}_{kj}} = \left[ \begin{array}{l}
{{\bf{a}}_{kj}}{\bf{a}}_{kj}^{\mathop{\rm H}\nolimits} ,{{\bf{a}}_{kj}}b_{kj}^ * \\
{b_{kj}}{\bf{a}}_{kj}^{\mathop{\rm H}\nolimits} ,\;\;\;0
\end{array} \right]$.

When the auxiliary variables ${\boldsymbol{\gamma}}$ and ${\boldsymbol{\varepsilon }}$ are both fixed, problem (22) is a convex problem, we can solve it via Mosek. After getting the optimized ${\bf Q}$, we can use the Gaussian randomization on ${\bf Q}$ to get a feasible rank-one solution ${\bf v}$ for problem (15) \cite{Wu18}. Therefore, by alternately solving problem (11) and problem (22), we can obtain a suboptimal solution ${\bf W}$ and ${\bf v}$ to the problem (9), the specific algorithm is presented in ${\bf{Algorithm\:1}}$.
Based on \cite{Wang14}, the complexity of ${\bf{Algorithm\:1}}$ is approximated as $\mathcal{O}\left( {{\rm{ite}}{{\rm{r}}_1}\sqrt {M + K} {M^3}{K^4} + {\rm{ite}}{{\rm{r}}_2}{K^{1.5}}{I^{6.5}}} \right)$,where ${\rm iter}_1$ and ${\rm iter}_2$ represent the number of iterations to solve problem (11) and problem (22), respectively.
\vspace{-0.2cm}
\begin{algorithm}
\caption{Beamforming and Phase shift for problem (9)}
\begin{algorithmic}[1]
\STATE Initialize ${\bf{W}}^{0}$ and ${\bf{v}}^{0}$ to a feasible value and set $t=0$.
\STATE For given ${\bf{v}}^{t}$, update ${\bf y}$ and $z$ by \eqref{12} and \eqref{13}.
\STATE Update ${\bf{W}}$ by solving problem (11) for fixed ${\bf y}$ and $z$.
\STATE Repeat steps 2-3 until EE does not increase and output the optimized beamforming ${\bf{W}}^{t+1}$.
\STATE For given ${\bf{W}}^{t+1}$, update ${\boldsymbol{\gamma}}$ and ${\boldsymbol{\varepsilon }}$ by \eqref{17} and \eqref{18}.
\STATE Update ${\bf{v}}$ by solving the problem (22) together with Gaussian randomization for fixed ${\boldsymbol{\gamma}}$ and ${\boldsymbol{\varepsilon }}$.
\STATE Repeat steps 5-6 until EE does not increase and output the optimized phase shifts ${\bf{v}}^{t+1}$.
\STATE If EE does not increase or $t=T$, the iteration process is terminated. Otherwise, set $t=t+1$ and go to step 2.
\end{algorithmic}
\end{algorithm}

\begin{figure*}
\setlength{\abovecaptionskip}{0.cm}
\setlength{\belowcaptionskip}{-0.cm}
\centering
\includegraphics[scale=0.6]{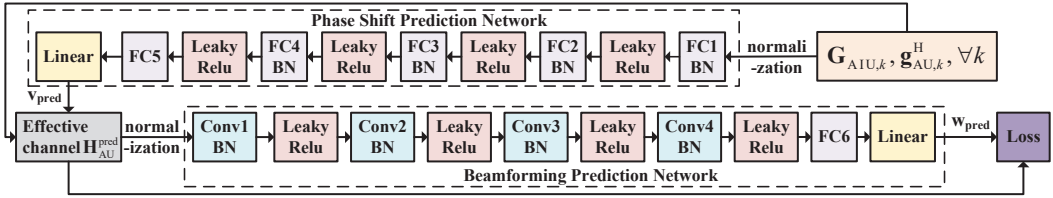}
\caption{An overview of the proposed network.}\label{fig2}
\end{figure*}
\vspace{-0.4cm}
\section{\small UNSUPERVISED LEARNING BASED APPROACH}
Unlike the time-consuming iterative optimization algorithm in section III, this section propose an unsupervised learning based approach to tackle problem (9), where this approach can quickly output the optimal $\bf W$ and $\bf v$ by inputting the channel state information of the corresponding scenario to the network trained by unsupervised learning manner.
\subsection{Network Architecture and Training}
The detailed network architecture and implementation are illustrated in Fig. 2. Based on problem (9), it is intuitive to use the real and imaginary parts of channels ${{\bf{G}}_{{\mathop{\rm AIU}\nolimits} ,k}}$ and ${{\bf{g}}_{{\mathop{\rm AU}\nolimits} ,k}^{\rm H}}$, $\forall k \in \mathcal{K}$ as the input of the network, where the total number of features is $2MIK + 2MK$. After these features are fed into the network, they will be normalized and fed into the phase shift prediction network. In the phase shift prediction network, we use five fully-connected (FC) layers, where the first four FC layers has $2MIK + 2MK$ neurons and the last layer has $I$ neurons, respectively. For the first four FC layers, each of them is first followed by the batch normalization (BN) layer and then followed by the Leaky rectified linear unit (LeakyRelu) activation layer. For FC5 layer, it uses a linear unit to output the phase shift prediction ${{\bf{v}}_{{\mathop{\rm pred}\nolimits} }} = \cos \left( {{{{\boldsymbol{\theta } }}_{{\mathop{\rm pred}\nolimits} }}} \right) + j\sin \left( {{{\boldsymbol{\theta }}_{{\mathop{\rm pred}\nolimits} }}} \right)$ based on the predicted real vector ${{\boldsymbol{\theta }}_{{\mathop{\rm pred}\nolimits} }}\in {\mathbb{R}^{I}}$. Based on ${{\bf{v}}_{{\mathop{\rm pred}\nolimits} }}$, optimal effective channel ${\bf{H}}_{{\mathop{\rm AU}\nolimits} }^{{\mathop{\rm pred}\nolimits} } = \left[ {{\bf{h}}_{{\mathop{\rm AU}\nolimits} ,1}^{\rm pred} ; \cdots ;{\bf{h}}_{{\mathop{\rm AU}\nolimits} ,K}^{\rm pred} } \right]$ with ${\bf{h}}_{{\mathop{\rm AU}\nolimits} ,k}^{\rm pred}  = {{\bf{v}}^{\mathop{\rm H}\nolimits}_{\rm pred} }{{\bf{G}}_{{\mathop{\rm AIU}\nolimits} ,k}} + {{\bf{g}}_{{\mathop{\rm AU}\nolimits} ,k}^{\rm H}}$ can be formed and then the real and imaginary parts of ${\bf{H}}_{{\mathop{\rm AU}\nolimits} }^{{\mathop{\rm pred}\nolimits}}$ will be normalized and used as input to the beamforming prediction network. In the beamforming prediction network, convolutional neural network is designed with four convolutional (Conv) layers with 256, 256, 128, 64 filters of size $M \times K$ and one FC layer with $2MK$ neurons, where FC layer is followed by a Linear unit to output the real and imaginary parts of the beamforming prediction ${\bf W}_{\rm pred}$. In addition, the usage of BN layer and activation function in beamforming prediction network is similar to the phase shift prediction network.

During training, we apply the Adam optimizer to update weights and set the initial learning rate to $0.01$. Maximal epoch, batch size and early stopping of patience are set as 500, 1024 and 20, respectively. To speed up the convergence, when the validation loss does not decrease within 10 epochs, the learning rate will be reduced by a factor of 0.5 until the minimum learning rate threshold $10^{-6}$ is reached. Besides, for sufficient training and accurate evaluation of model, we generate $1.6 \times 10^5$, $4 \times 10^4$ and $10^3$ samples for training, validation and testing, respectively. Finally, it is worth noting that since unsupervised learning does not require labels, it greatly reduces the cost of obtaining training samples.

\subsection{Loss Function}
Given that the goal of problem (9) is to maximize EE under some constraints of (9b), (9c) and (9d), we apply the following loss function for model training as
\setcounter{equation}{22}
\begin{equation}\label{23}
{\mathop{\rm Loss}\nolimits}  = \frac{1}{T}\sum\limits_{t = 1}^T {\left( \begin{array}{l}
 - {\mathop{\rm EE}\nolimits}  + \sum\nolimits_{k = 1}^K {{\beta _{\rm{1}}}{{\left[ {{R_{\min }-{R_k}}} \right]}^ + }} \\
 + \sum\nolimits_{m = 1}^M {{\beta _{\rm{2}}}{{\left[ {{\left\| {{{{\bf{\bar w}}}_m}} \right\|^2}  - {P_{\max }}} \right]}^ + }}
\end{array} \right)},
\end{equation}
where $T$ is the mini-batch size. ${\beta _1}$ and ${\beta _2}$ are the penalty coefficient for the rate constraint of (9b) and the power constraint of (9c), where each penalty term will have a positive value only if the corresponding constraints are violated, enforcing the training process towards satisfying the given requirement. In addition, the constraint of (9d) is not added to loss function because it is already implemented in the process of converting the real vector ${{\boldsymbol{\theta }}_{{\mathop{\rm pred}\nolimits} }}$ to ${{\bf{v}}_{{\mathop{\rm pred}\nolimits} }}$. Therefore, by choosing appropriate ${\beta _1}$ and ${\beta _2}$, loss function can be used to solve problem (9) efficiently. Finally, based on \cite{He15}, the complexity of unsupervised learning based approach is approximated as $\mathcal{O}\left( {{M^2}{K^2}\left( {16{I^2} + {1.07*{10}^5}} \right)} \right)$.

\section{Numerical Results}
In this section, simulation results are provided to validate the EE performance of the iterative optimization algorithm and the unsupervised learning based approach. First, we consider a simulation setup, where $M=13$ APs and $L=2$ IRSs configured with $N = 50$ reflecting elements provide services to $K=3$ single-antenna users. Besides, the $m$-th AP is located at $\left( {10\left( {m - 1} \right)\;{\mathop{\rm m}\nolimits} , - 40\;{\mathop{\rm m}\nolimits} ,5\;{\mathop{\rm m}\nolimits} } \right), m \in \mathcal{M}$, and the IRSs are located at $\left( {40\;{\mathop{\rm m}\nolimits} ,10\;{\mathop{\rm m}\nolimits} ,10\;{\mathop{\rm m}\nolimits} } \right)$ and $\left( {80\;{\mathop{\rm m}\nolimits} ,10\;{\mathop{\rm m}\nolimits} ,10\;{\mathop{\rm m}\nolimits} } \right)$, respectively. Finally, the users with a height of 1.65 m are located in a uniform circular area with a radius of 2 m, and the center of the region can move along the line between $\left( {0\;{\mathop{\rm m}\nolimits} ,0\;{\mathop{\rm m}\nolimits}} \right)$ and $\left( {120\;{\mathop{\rm m}\nolimits} ,0\;{\mathop{\rm m}\nolimits}} \right)$.

For all channels in simulation setup, we apply a Rician fading channel modeling approach similar to reference \cite{Jin22}. For example, we assume that the path loss is 30 dB at the reference distance 1 m. The path loss exponent of the AP-IRS link, the IRS-user link and the AP-user link are set as 2.2, 2.8 and 3.5, respectively. The Rician factor of the AP-IRS link, the IRS-user link and the AP-user link are set as 10 dB, 5 dB and $- \infty$ dB, respectively. Besides, other simulation parameters are given as the maximum transmit power of each AP ${P_{\max }} = 0\;{\mathop{\rm dB}\nolimits} $, the minimum rate requirement of each user ${R_{\min }} = 1\;{\rm bit/s/Hz}$, the circuit power of each AP ${P_{{\mathop{\rm AP}\nolimits} }} = 10\;{\mathop{\rm dBm}\nolimits}$, the circuit power of each user ${P_{{\mathop{\rm User}\nolimits} }} = 10\;{\mathop{\rm dBm}\nolimits}$, the circuit power of each IRS element ${P_{{\mathop{\rm IRS}\nolimits} }} = 0\;{\mathop{\rm dBm}\nolimits}$, the bandwidth $B = 1\;{\mathop{\rm MHz}\nolimits}$, the noise power ${\sigma ^2} =  - 60\;{\mathop{\rm dBm}\nolimits}$, the power amplifier efficiency $\upsilon=0.8$, and the penalty coefficient ${\beta _1} = {\beta _2} = 50$, respectively. Finally, all operations are executed by software (Python 3.6, TensorFlow 2.6.0 and Mosek 9.3.20) and hardware (two Xeon Platinum 8368 CPUs, a GeForce RTX 3090 GPU and 512 GB RAM).

\begin{figure}[thb!]
\setlength{\abovecaptionskip}{0.cm}
\setlength{\belowcaptionskip}{-0.cm}
\centering
\includegraphics[scale=0.33]{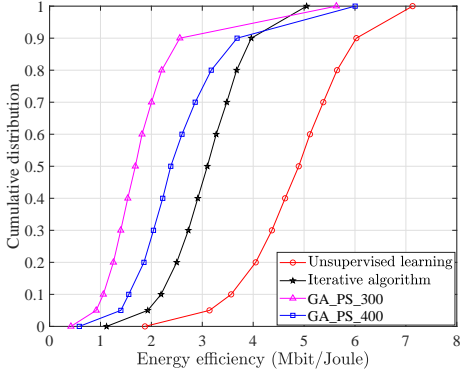}
\caption{Cumulative distribution of the energy efficiency.}\label{fig3}
\end{figure}

\begin{figure}[thb!]
\setlength{\abovecaptionskip}{0.cm}
\setlength{\belowcaptionskip}{-0.cm}
\centering
\includegraphics[scale=0.67]{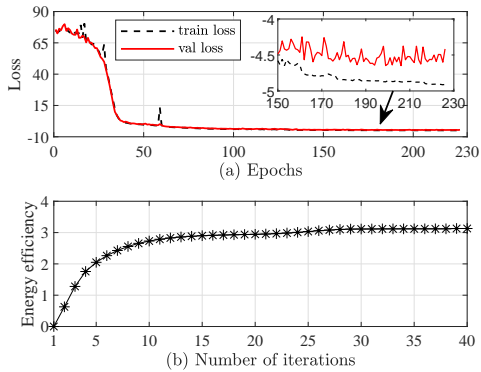}
\caption{The loss of unsupervised learning based approach on the epochs and the EE of iterative optimization algorithm on the number of iterations.}\label{fig4}
\end{figure}

To validate the efficiency of our proposed scheme, Fig. 3 compares the cumulative distribution functions of EE for the iterative optimization algorithm, the unsupervised learning based approach and the GA, where the population size (PS) and the maximum number of generations play a key role in the performance of GA. In our simulations, we set the maximum number of generations to 20000 and set the PS to $\left\{ {300,400} \right\}$, respectively. In Fig. 3, we can observe that the $95\%-{\rm{likely}}$ EE by the unsupervised learning based approach, the iterative optimization algorithm, the GA with PS 300 and PS 400 are about 3.14 Mbit/Joule, 1.92 Mbit/Joule, 0.91 Mbit/Joule and 1.39 Mbit/Joule, respectively. Besides, the average running time of these schemes are given as 0.001 s, 6.592 s, 18.281 s and 21.977 s, respectively. Given the highest EE performance and the lowest running time, the unsupervised learning based approach is regarded as an optimal scheme compared to other schemes. In addition, we can also observe that for the $95\%-{\rm{likely}}$ EE, the iterative optimization algorithm only consumes about $30\%$ of the running time to reach 2.1 (or 1.38) times EE performance of GA with PS 300 (or 400), respectively. Since the unsupervised learning based approach takes a lot of time to train a perfect DNN, the iterative optimization algorithm is also considered a competitive approach in scenarios where training time is limited.

An exemplary training process is illustrated in Fig. 4(a). As can be readily observed, our proposed DNN architecture can approach convergence after about 100 epochs. This verifies the feasibility of the DNN architecture. In addition, the impact of iteration number on the EE of the iterative optimization algorithm is illustrated in Fig. 4(b). It can be observed that as the number of iterations increases, the EE can approach convergence after 20 iterations. This also shows that the iterative optimization algorithm has good convergence.
\vspace{-0.15cm}
\section{Conclusions}
For the IRS-aided cell-free massive MIMO system, we proposed an iterative optimization algorithm and an unsupervised learning based approach to tackle EE maximization problem, respectively. The simulation results show that, benefiting from the universal approximation properties of the DNN, the unsupervised learning based approach exceeds the iterative optimization algorithm and GA in terms of EE and running time. This shows that the unsupervised learning based approach is ideal for solving EE maximization problem. In addition, when the training time is limited, the iterative optimization algorithm performs relatively well in terms of EE, running time and convergence, indicating that it is also an excellent scheme.


\vspace{-0.15cm}
\begin{thebibliography}{1}
\bibitem{Ngo17}  H. Q. Ngo, A. Ashikhmin, H. Yang, E. G. Larsson, and T. L. Marzetta, ``Cell-free massive MIMO versus small cells,'' {\em IEEE Trans. Wireless Commun.}, vol.~16, no.~3, pp.~1834--1850, Mar. 2017.
\bibitem{Wu19} Q. Wu and R. Zhang, ``Intelligent reflecting surface enhanced wireless network via joint active and passive beamforming,'' {\em IEEE Trans. Wireless Commun.}, vol.~18, no.~11, pp.~5394--5409, Nov. 2019.
\bibitem{Wu18}  Q. Wu and R. Zhang, ``Intelligent reflecting surface enhanced wireless network: Joint active and passive beamforming design,'' in {\em Proc. IEEE Global Commun. Conf. (GLOBECOM)}, Dec. 2018, pp. 1-6.
\bibitem{Jin22} S.-N. Jin, D.-W. Yue, and H. H. Nguyen, ``RIS-aided cell-free massive MIMO system: Joint design of transmit beamforming and phase shifts,'' {\em IEEE Syst. J.}, early access, Aug. 8, 2022, doi: 10.1109/JSYST.2022.3194259.
\bibitem{ZZhang21}  Z. Zhang and L. Dai, ``A joint precoding framework for wideband reconfigurable intelligent surface-aided cell-free network,'' {\em IEEE Trans. Signal Process.}, vol.~69, pp.~4085--4101, Jun. 2021.
\bibitem{Zhang21}  Y. Zhang {\em et al.}, ``Beyond cell-free MIMO: Energy efficient reconfigurable intelligent surface aided cell-free MIMO communications,'' {\em IEEE Trans. Cogn. Commun. Netw.}, vol.~7, no.~2, pp.~412--426, Jun. 2021.
\bibitem{Le21}  Q. N. Le, V. -D. Nguyen, O. A. Dobre, and R. Zhao, ``Energy efficiency maximization in RIS-aided cell-free network with limited backhaul,'' {\em IEEE Commun. Lett.}, vol.~25, no.~6, pp.~1974--1978, Jun. 2021.
\bibitem{Siddiqi22}  M. Z. Siddiqi, R. Mackenzie, M. Hao, and T. Mir, ``On energy efficiency of wideband RIS-aided cell-free network,'' {\em IEEE Access}, vol.~10, pp.~19742--19752, 2022.
\bibitem{Liu21} K. Liu and Z. Zhang, ``On the energy-efficiency fairness of reconfigurable intelligent surface-aided cell-free network,'' in {\em Proc. IEEE 93rd Veh. Technol. Conf. (VTC2021-Spring)}, Apr. 2021, pp.~1--6.
\bibitem{Song21}  H. Song, M. Zhang, J. Gao, and C. Zhong, ``Unsupervised learning-based joint active and passive beamforming design for reconfigurable intelligent surfaces aided wireless networks,'' {\em IEEE Commun. Lett.}, vol.~25, no.~3, pp.~892--896, Mar. 2021.
\bibitem{Huang20}  C. Huang, R. Mo, and C. Yuen, ``Reconfigurable intelligent surface assisted multiuser MISO systems exploiting deep reinforcement learning,'' {\em IEEE J. Sel. Areas Commun.}, vol.~38, no.~8, pp.~1839--1850, Aug. 2020.
\bibitem{Wang20}  Z. Wang, L. Liu, and S. Cui, ``Channel estimation for intelligent reflecting surface assisted multiuser communications: Framework, algorithms, and analysis,'' {\em IEEE Trans. Wireless Commun.}, vol.~19, no.~10, pp.~6607--6620, Oct. 2020.
\bibitem{Shen18} K. Shen and W. Yu, ``Fractional programming for communication systems-Part I: Power control and beamforming,'' {\em IEEE Trans. Signal Process.}, vol.~66, no.~10, pp.~2616--2630, May. 2018.
\bibitem{Mosek} {\em Mosek Aps}, MOSEK Inc., Copenhagen, Denmark, 2014. [Online]. Available: www.mosek.com
\bibitem{KShen18} K. Shen and W. Yu, ``Fractional programming for communication systems-Part II: Uplink scheduling via matching,'' {\em IEEE Trans. Signal Process.}, vol.~66, no.~10, pp.~2631--2644, May. 2018.
\bibitem{Wang14} K.-Y. Wang, A. M.-C. So, T.-H. Chang, W.-K. Ma, and C.-Y. Chi, ``Outage constrained robust transmit optimization for multiuser MISO downlinks: Tractable approximations by conic optimization,'' {\em IEEE Trans. Signal Process.}, vol.~62, no.~21, pp.~5690--5705, Nov. 2014.
\bibitem{He15} K. He and J. Sun, ``Convolutional neural networks at constrained time cost,'' in {\em Proc. IEEE Conf. Comput. Vis. Pattern Recognit. (CVPR)}, Jun. 2015, pp.~5353--5360.
\end{thebibliography}
\end{document}